\documentclass[letterpaper,twocolumn,10pt]{article}
\usepackage{usenix2019_v3}

\usepackage{tikz}
\usepackage{amsmath}

\usepackage{graphicx}
\usepackage{xcolor}
\usepackage{comment}
\usepackage{subcaption}
\usepackage[T1]{fontenc}
\usepackage{booktabs}
\usepackage{makecell}

\newcommand*\emptycirc[1][1ex]{\tikz\draw (0,0) circle (#1);} 
\newcommand*\halfcirc[1][1ex]{%
  \begin{tikzpicture}
  \draw[fill] (0,0)-- (90:#1) arc (90:270:#1) -- cycle ;
  \draw (0,0) circle (#1);
  \end{tikzpicture}}
\newcommand*\fullcirc[1][1ex]{\tikz\fill (0,0) circle (#1);}

\microtypecontext{spacing=nonfrench}
\usepackage[switch]{lineno}

\begin{document}
\date{}

\title{\Large \bf SoK: State of the time: On Trustworthiness of Digital Clocks}

\author{
{\rm Adeel Nasrullah}\\
University of Massachusetts Amherst
\and
{\rm Fatima M. Anwar}\\
University of Massachusetts Amherst
}

\maketitle

\begin{abstract}
Despite the critical role of timing infrastructure in enabling essential services—from public key infrastructure and smart grids to autonomous navigation and high-frequency trading—modern timing stacks remain highly vulnerable to malicious attacks. These threats emerge due to several reasons, including inadequate security mechanisms, the timing architecture's unique vulnerability to delays, and implementation issues. In this paper, we aim to obtain a holistic understanding of the issues that make the timing stacks vulnerable to adversarial manipulations, what the challenges are in securing them, and what solutions can be borrowed from the research community to address them. To this end, we perform a systematic analysis of the security vulnerabilities of the timing stack. In doing so, we discover new attack surfaces, i.e., \textit{physical timing components and on-device timekeeping}, which are often overlooked by existing research that predominantly studies the security of time synchronization protocols. We also show that the emerging trusted timing architectures are flawed \& risk compromising wider system security, and propose an alternative design using \emph{hardware-software co-design}.
\end{abstract}

\section{Introduction}\label{sec:introduction}
An accurate perception of time is indispensable in modern digital ecosystems. Critical applications such as Public Key Infrastructure (PKI)\cite{time-stack-abouttime}, streaming authentication protocols such as TESLA~\cite{tesla-cryptography}, smart grid operations\cite{intro-tech-report-smart-grid, intro-smart-grid-pmu-attack}, autonomous vehicle perception systems\cite{hardware-chronos-slam-attack}, and high-frequency trading\cite{intro-high-frequency-trading} depend on precise timing for their proper functionality. Further, the absence of accurate time synchronization endangers \textit{Time Difference of Arrival} (TDoA)-based applications, which are vital for electronic warfare\cite{intro-electronic-warfare}, vehicular triangulation\cite{intro-vehicle}, sonar operations in marine environments\cite{intro-sonar}, environmental conservation\cite{intro-gunshot-localization}, and indoor localization\cite{intro-indoor-localization}. Disrupting these cyber-physical systems (CPS), by exploiting their timing stacks, poses a serious threat to society.

Despite the pivotal role of timing infrastructure, it remains susceptible to adversarial manipulations. NTP~\cite{ntpv4-rfc}, the predominant time-sync protocol on the internet, could face continental-scale disruptions due to a few malicious servers within its pool~\cite{shark-ntp-pool}. PTP~\cite{ptp-std-doc}, essential for high-precision synchronization, is vulnerable to frequency manipulations by compromised nodes despite using dual authentication~\cite{net-sync-ptp-covert-channel}. GPS, integral to critical infrastructure, can be easily spoofed~\cite{gps-spoofing-fundamentals}. Furthermore, recently developed high-precision time-sync protocols, e.g., Huygens~\cite{huygens} and Sundial~\cite{sundial}, as well as the IoT framework \textit{Matter}'s timing stack~\cite{matter}, do not adequately address their security. The prevalence of security vulnerabilities in existing and emerging time stacks warrants a systematic examination of their issues.

In this paper, we present the first systematization of knowledge for time security in a typical CPS. While existing research on this topic has made important contributions, it has been limited to examining single protocols (e.g., NTP~\cite{ntp-replay-drop-attack, shark-ntp-pool, theory-nts-specs}, PTP~\cite{ptp-futile-encryption, net-sync-ptp-sec, net-sync-ptp-covert-channel}, and GPS~\cite{gps-spoofing-fundamentals, gps-anti-jamming-post-correlation, gps-anti-jamming-post-wavelet, gps-anti-spoofing-static} etc.) or specific types of attacks (e.g., delay attacks~\cite{ptp-futile-encryption, multi-path-game-theory}). In contrast, we propose the idea of a timing framework and provide a holistic view of time stack security; we analyze the vulnerabilities of \textit{physical timing components} (e.g., quartz crystals), \textit{software-based clocks}, and the \textit{time-sync protocols}. Utilizing this timing framework, we highlight existing and emerging attack surfaces, gain insights into the extent and scope of proposed countermeasures, and identify open research problems.

Leveraging our framework, we discover previously uncharted attack surfaces threatening the timing stack of CPS, and present case studies that underscore the gravity of these threats. Through our study, we demonstrate that these attacks surfaces, constituting \textit{side-channel attacks on physical timing components}, and \textit{risks posed by privileged software to the integrity of system time}, are insufficiently addressed by the current body of timing security research. This oversight often results in conflating timing stack security with mere time synchronization issues, predominantly attributing threats to \textit{network-based attackers} (e.g., DNS cache poisoning on NTP~\cite{shark-ntp-pool}, delay attacks on PTP~\cite{ptp-futile-encryption}, and GPS spoofing~\cite{intro-attacks-critical}). While some recent research has focused on securing time within Trusted Execution Environments (TEEs) to guard against privileged software adversary, our analysis reveals that this area is fraught with unresolved challenges. Our work fills this gap through a comprehensive systematization of security vulnerabilities across the timing stack's hardware, software, and network dimensions.

We also show that research methodologies used for building secure timing services can be classified into two categories: \textit{system-based} approach, which utilizes design mechanisms like NTP's message authentication~\cite{ntpv4-rfc} to deter attacks, and \textit{theoretical} approach, applying mathematical tools (e.g., theorem proving, model checking, game theory) for the security analysis of timing protocols. Despite their significant contributions to timing stack security, very few works successfully utilize both strategies together. Their limited interaction is exemplified by the formal analysis of NTS~\footnote{NTS is a secure version of NTP.} by Tiechel et al.~\cite{theory-nts-formal-analysis}, which falls short of analyzing full specifications due to the analysis tool's inability to model time and clocks. Nevertheless, we evaluate the two approaches to determine their coverage of the timing stack's security issues and contrast their contributions.

Moreover, we identify a significant risk to CPS security posed by state-of-the-art trusted timing services like Timeseal~\cite{time-stack-timeseal} and T3E~\cite{trusted-time-t3e}. These solutions provide TEE~\footnote{Trusted Execution Environment.}-confined trusted timestamps to user applications, requiring them to execute inside the TEE. Given that such applications may come from untrusted sources, they pose a threat to sensitive code and data inside the TEE. As a result, these solutions inadvertently increase system-wide security risks. Drawing from our timing stack security analysis, we provide recommendations for an alternative design that aligns with the goal of overall system security. Additionally, we highlight future research directions to advance this domain.

In summary, we make the following contributions: (1) We conduct the first systematic study of timing stack security in light of the proposed timing framework. (2) We identify previously overlooked attack surfaces through case studies. (3) We propose an intuitive taxonomy for categorizing timing stack vulnerabilities and analyze relevant literature. (4) We categorize timing stack solutions into system-based and theoretical, highlighting their unique contributions and limitations. (5) Finally, we identify open research challenges related to the timing stack's security.

\section{Background} \label{background}
In this section, we provide a brief introduction to the timing stacks in contemporary digital systems. Additionally, we describe the types of timing attacks and present case studies demonstrating attacks on the timing stack.

\subsection{Timing Stack Basics}\label{subsec:local-clocks}
\begin{figure}[h]
        \centering           
        \includegraphics[width=0.5\columnwidth]{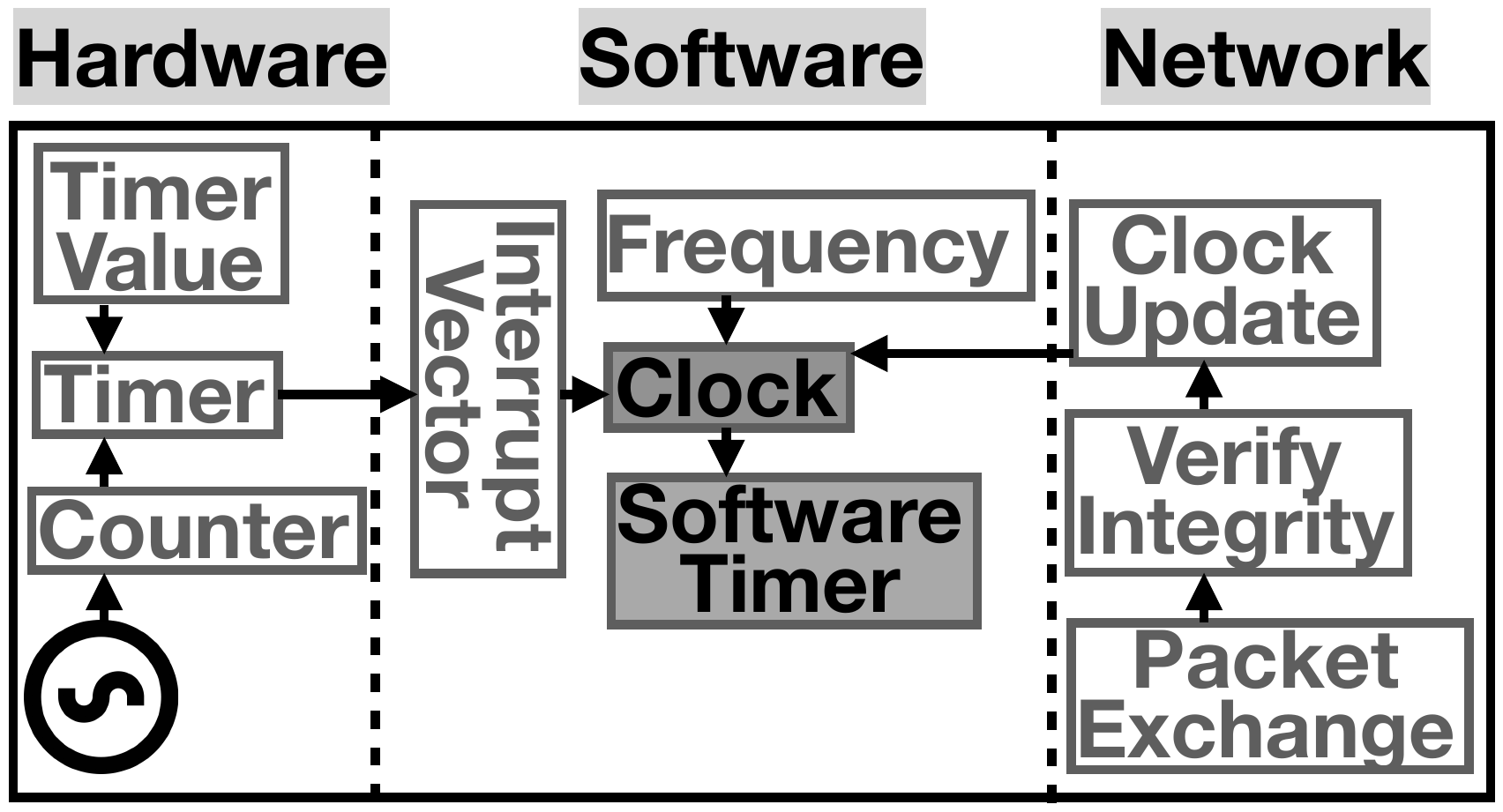}
        \caption{Time stack in a modern CPS}
        \label{fig:time-stack-example}
\end{figure}

\noindent\textbf{Measuring Time.}
The foundation of any timing stack is its timing source, which emits a recurring signal at fixed intervals. This \textit{clock signal} is essential for digital systems to measure time accurately. Figure~\ref{fig:time-stack-example} illustrates a typical Cyber-Physical System (CPS) timing stack, where the time source is integrated into the physical hardware, accompanied by \textit{time measurement} components such as counters and timer registers. The time source, often a quartz crystal, produces an analog signal that feeds into these components. The counter records the number of \textit{clock periods (cycles)} since the system was powered on, whereas the timer registers are designed to initiate an interrupt after a predetermined number of cycles. System software utilizes these elements to maintain \textit{time} and distribute it to user applications.

\noindent\textbf{Keeping Time.} Hardware counters and timers measure time in clock cycles, a unit that varies across systems due to differences in time source frequencies. Furthermore, this measurement begins at the system's power-up, an arbitrary starting point. Thus, time obtained from the hardware counters and timers is non-standardized. System software establishes a \textit{local clock} by converting these clock cycles into standard wall clock time using the clock signal frequency and network-derived current time (Figure~\ref{fig:time-stack-example}). This local clock is updated at regular intervals using a recurring timer interrupt called \textit{system tick}. When a timestamp is needed between two consecutive system ticks, the system software reads the current value of the processor \textit{counter} to determine the time elapsed since the last \textit{system tick} and adds it to the time recorded at the last tick to compute the current time~\cite{linux-hrtimers}. Similarly, it also maintains a \textit{software timer} providing a standardized interface to the user applications.

\begin{figure}[t]
    \centering
    \begin{subfigure}{0.28\columnwidth}
        \centering   
     \includegraphics[width=\linewidth]{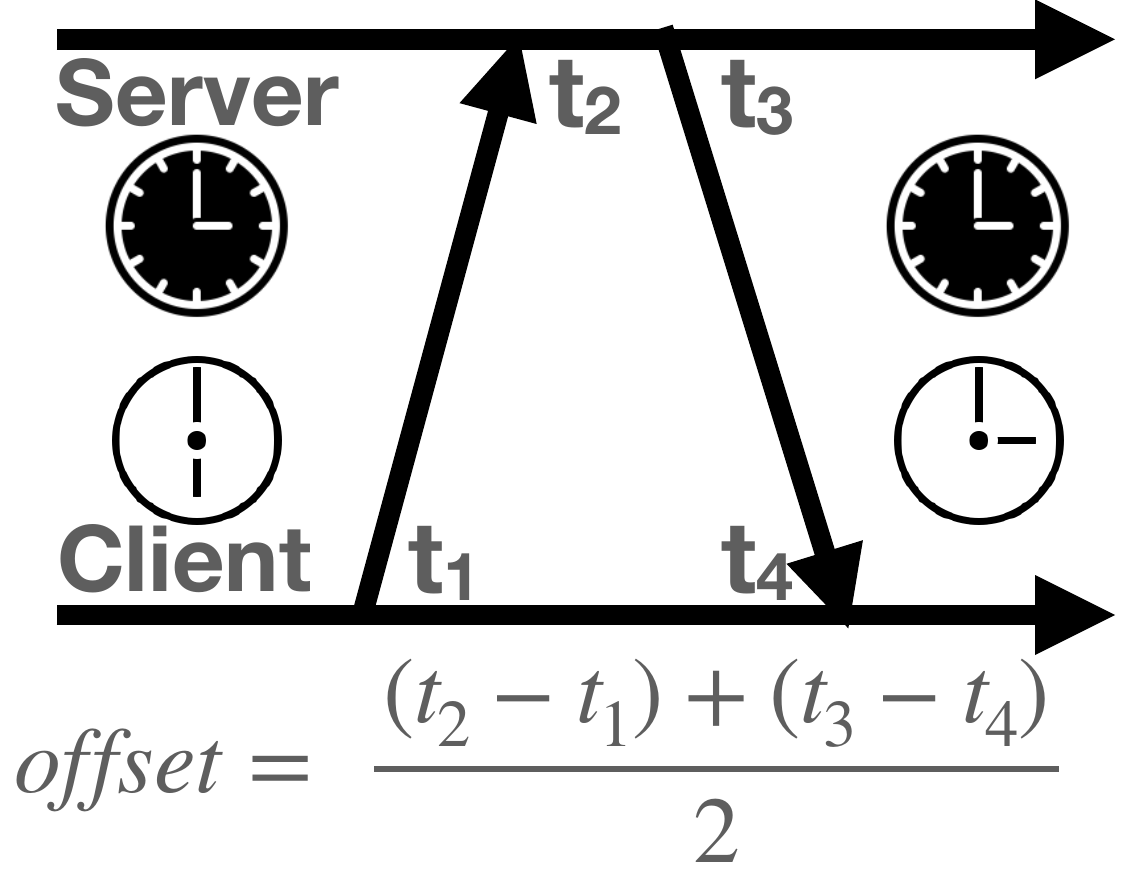}
        \caption{}
        \label{fig:two-way}
    \end{subfigure}
    \hspace{2em}
    \begin{subfigure}{0.35\columnwidth}
        \centering 
    \includegraphics[width=\linewidth]{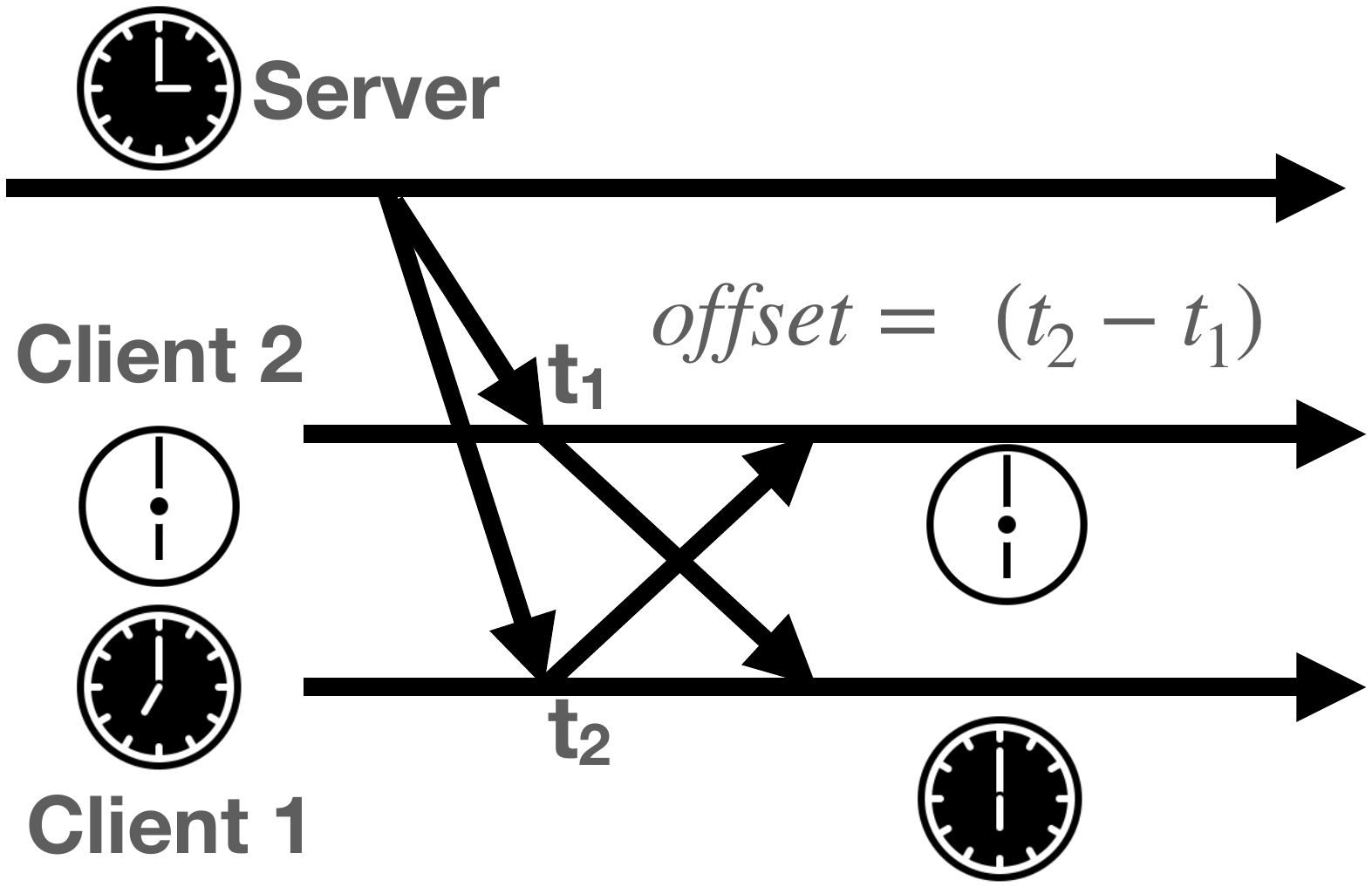}
        \caption{}
        \label{fig:one-way}
    \end{subfigure}
    \caption{(a) Two-way time synchronization. (b) One-way time synchronization.}
    \label{fig:sync-paradigms}
\end{figure}

\noindent\textbf{Time Synchronization.} Time source frequency variation, influenced by environmental factors like temperature, causes \textit{local clocks} to drift from the actual time~\cite{graham-clock-sync}. Time synchronization services correct this by estimating the local clock's deviation from a network-provided reference clock, using packet exchanges and integrity checks to securely adjust the local clock (Figure~\ref{fig:time-stack-example}). Time-sync protocols fall into two categories: (1) \textit{Two-way Time-Sync} utilizes bidirectional message exchange to compute offset and skew~\footnote{Offset and skew refer to the baseline time difference between two clocks and the rate at which this difference grows, respectively.}. These parameters are utilized to align local clocks with an external reference (Figure~\ref{fig:two-way}).(2)  \textit{One-way Time-Sync} relies on one-way broadcasts from servers to clients, as used by GPS~\cite{gps-spoofing-fundamentals} and some sensor network protocols~\cite{Elson2003RBS} (Figure~\ref{fig:one-way}).

\subsection{Temporal Manipulations}\label{subsec:attack-types}
\begin{figure}[ht]
    \small
    \centering
    \begin{subfigure}{0.48\columnwidth}
        \centering
        \includegraphics[width=\columnwidth]{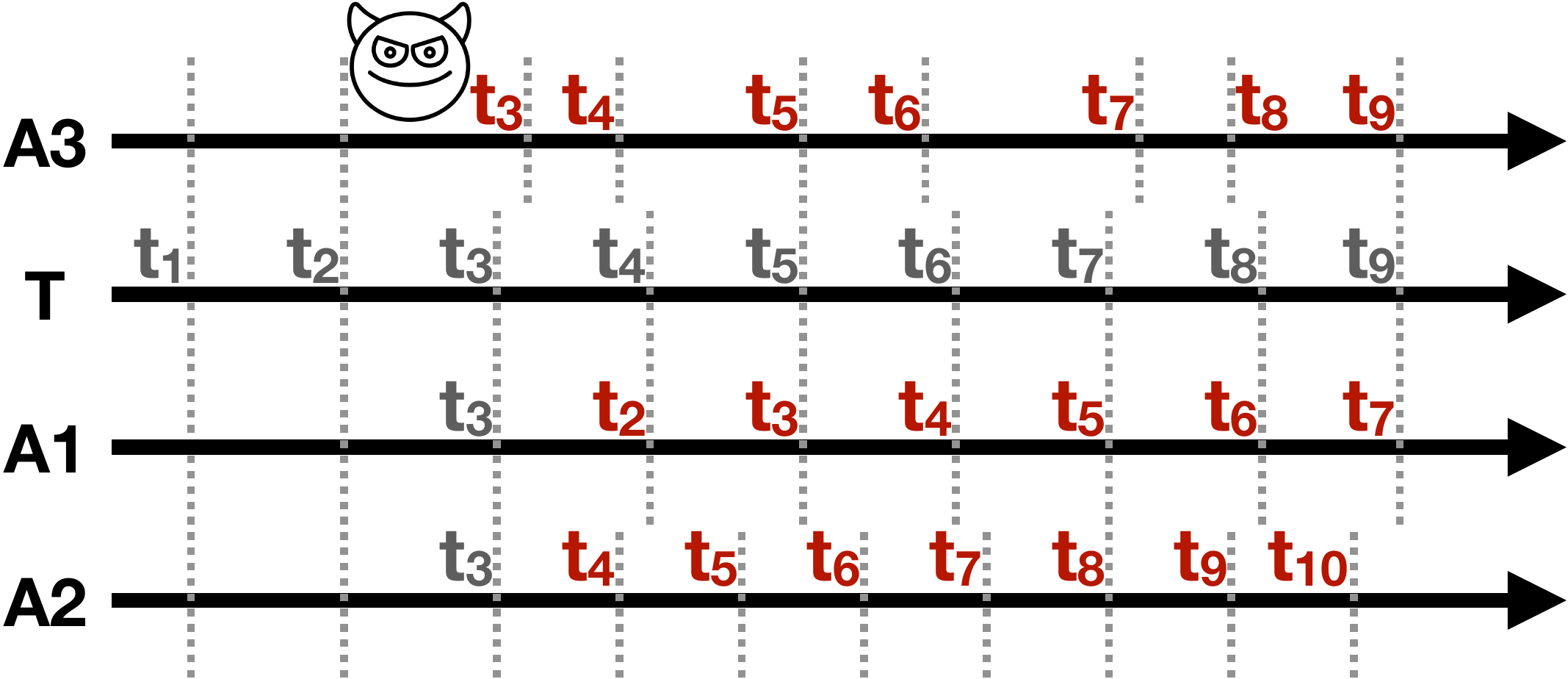}
        \caption{}
        \label{fig:attack-types}
    \end{subfigure}
    \hspace{0.2em}
    \begin{subfigure}{0.48\columnwidth}
        \centering
        \includegraphics[width=\columnwidth]{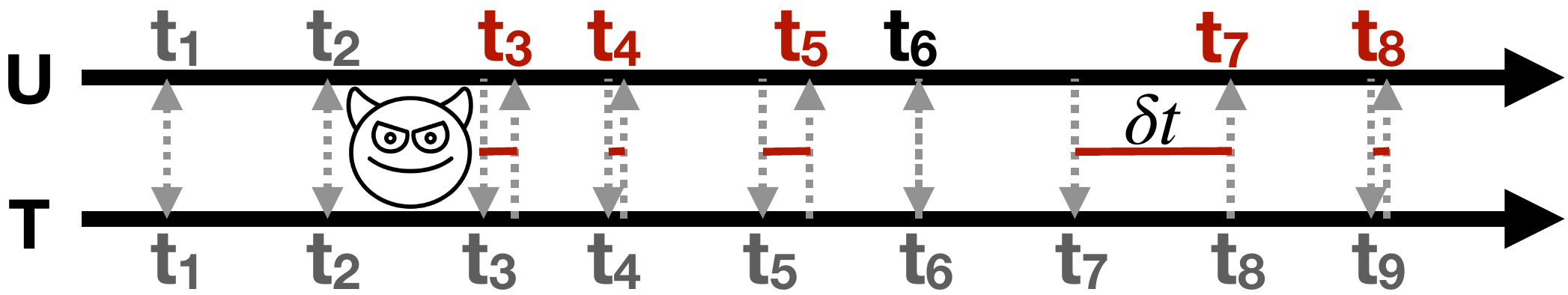}
        \caption{}
        \label{fig:delay-attack}
    \end{subfigure}
    \caption{(a) \textbf{\texttt{T}} represents the system time under normal conditions. \textbf{\texttt{A1}}, \textbf{\texttt{A2}}, and \textbf{\texttt{A3}} illustrate the time under time travel, warping, and uncertainty attacks, respectively. (b) Agent \textbf{\texttt{U}} obtains its time from another agent, \textbf{\texttt{T}}, assuming instantaneous time transfer. After time $t_2$, an attacker takes control of agent $T$ and delays the time transfer by a duration $0 \leq \delta t \leq \infty$.} 
    \label{fig:timing-attack-types}
\end{figure}

Malicious adversaries may seek to attack the timing stacks of cyber-physical systems, just as they target other components of the digital infrastructure. Such attacks aim to violate one or more of an ideal time stack's properties: 
(\textbf{\texttt{P1}}) the \textit{local clock} is monotonic, i.e., time always moves forward, (\textbf{\texttt{P2}}) it maintains a constant frequency, i.e., moves at a fixed rate, (\textbf{\texttt{P3}}) its frequency matches that of a reference clock, and (\textbf{\texttt{P4}}) it provides time relative to the epoch\footnote{A fixed date and time (Jan 1, 1970) used as a reference from which a system measures time.} also used by the reference clock. Figure~\ref{fig:attack-types} illustrates different forms of timing attacks that result from the violation of these four properties.

\noindent\textbf{(\texttt{A1}) Time Travel:} The local clock travels back or forward in time, violating \textbf{\texttt{P1}} and \textbf{\texttt{P4}}.

\noindent\textbf{(\texttt{A2}) Time Warping:} The local clock moves slower or faster relative to the reference, distorting the system's perception of time. This violates \textbf{\texttt{P3}} and, consequently, the discrepancy between local and the reference clock grows over time.

\noindent\textbf{(\texttt{A3}) Increased Uncertainty:} This attack targets the precision of the local clock (\textbf{\texttt{P2}}) while generally maintaining \textbf{\texttt{P3}} in the long-term. It reduces the effective temporal resolution of the local clock; for example, a clock that should provide time accurate to a millisecond is now only reliable to the second.

\subsection{Timing Attacks and CPS}\label{subsec:case-studies}
We demonstrate the vulnerability of CPS to timing attacks through various case studies, illustrating how adversaries exploit timing stack weaknesses to launch the attacks, described above.

\noindent\textbf{\texttt{C1.} No Trace Industrial Sabotage.} Industrial control systems are highly sensitive to temporal uncertainties~\cite{hardware-butterfly, hardware-polyrythem} that could be exploited by adversaries. An attacker can create temporal uncertainties just by exposing the control unit's quartz crystal (\textit{time source}) to lasers~\cite{redshift}. It would vary the crystal's frequency substantially, inducing \textit{time warping} (\textbf{\texttt{A2}}) to the control unit's \textit{local clock} and causing unstable behavior. An insider with \textit{physical access} to industrial equipment could launch such an attack without leaving any digital traces. 

\noindent\textbf{\texttt{C2.} Zero Knowledge Attack on AV-perception.} Autonomous vehicles (AV) utilize deep learning based multi-modal perception, relying on tightly synchronized sensor data~\cite{chen2019selectfusion}. Lack of synchronized inputs can destabilize these systems with potentially fatal consequences~\cite{hardware-chronos-slam-attack}. An attacker could introduce such de-synchronization by following these steps: i) maintain multiple copies of the \textit{local clock} on the victim AV, each moving at a different pace (\textbf{\texttt{A2}}) i.e. asynchronous clocks. ii) It then presents a different clock to each sensor subsystem disrupting their mutual synchronization and consequently causes the AV perception system to malfunction. This attack requires \textit{privileged access} to the AV system software, and identification of sensor subsystems. The former could be enabled by the privilege escalation vulnerabilities, that are discovered routinely (table~\ref{tab:cve-stats}), in the commodity system software. And the sensor subsystems are identified using the fact that they request repeated timestamps with a fixed interval between consecutive requests. Finally, we argue that this exploit lowers the cost of attacking AV perception because, in contrast to traditional attacks~\cite{hallyburton2022security, av-spoofing-attack}, it does not require any machine learning knowledge on part of the attacker nor does it require physical proximity to the victim.

\noindent\textbf{\texttt{C3.} Database Performance Degradation.} The consistency of database systems hinges on precise timekeeping~\cite{huygens}. To elaborate further, consider a database whose \textit{local clock} has an uncertainty of $\Delta t$ seconds. It receives two write requests at times $t_1$ and $t_2$, respectively, where $t_1-t_2 < \Delta t$. Due to its local clock's uncertainty, the database cannot determine the order in which two requests were issued. If it completes the two transactions, it risks losing data by committing write requests in wrong order. To avoid such issues many databases such as Google's spanner choose to process transaction only if they are temporally spaced by at least  $\Delta t$ (uncertainty in the \textit{local clock}'s time)~\cite{google-spanner}. An adversary with the \textit{privileged access} to the system software can manipulate the \textit{local clock} to increase uncertainty $\Delta t$ in its time \textbf{\texttt{A3}}. This will increase the wait times between consecutive transactions and severely degrade the database performance. In contrast, if a database does not wait out these timing uncertainties, such an attack would result in database inconsistencies.

\noindent\textbf{\texttt{C4.} Manipulating Smart Contract Systems.} Smart contracts require secure time synchronization between the client (the offeree) and the server (the offerer) devices. It establishes the order of the events such as contract offering, modifications and signing. This ordering is indispensable for these events' validity in the case of a litigation~\cite{smart-contract-tabellion}. Consider a scenario where an offeree may agree to unfavourable terms to secure the contract by out-competing rivals. However, before signing the contract, the offeree changes its device's system time to the past when the unfavorable clauses were yet to be added (\textbf{\texttt{A1}}). They can get away with this, if the offerer, having trust in the contract system, does not notice this discrepancy. In the case of a dispute, it enables the offeree to make a plausible case that they did not agree to contentious terms causing financial losses to the offerer. It is important to note that, if the smart contract system gets its time from the network, the offeree can still rewind the clock back, albeit using a more sophisticated attack (see section~\ref{sec:network-issues}). 

\noindent\textbf{\texttt{C5.} GPS manipulation example.} GPS is indispensable to the navigation systems in aviation, maritime trade, and public transport, among others. Its adversarial manipulation could lead to significant financial and human costs. Adversarial attacks on GPS exploit its dependence on signals broadcast by satellites with tightly synchronized clocks. GPS uses the relative delay in the reception of these signals to estimate its location and clock offset relative to the satellite clocks. The adversary captures and replays a legitimate satellite signal with a delay~\cite{gps-spoofing-fundamentals} (\textbf{\texttt{A3}}) to cause uncertainty in the location and time perceived by the receiver. Furthermore, a single antenna is sufficient for launching this attack against a victim, as shown by Tippenhauer et al.,~\cite{gps-spoofing-fundamentals}. The only constraint is that the attacker's signal power, as received by the victim, should be higher than the legitimate GPS signals. Alternatively, an adversary with \textit{physical access} to the GPS receiver could remove the antenna and attach a low-cost device that generates a fake GPS signal to the antenna input, achieving the same results.

It is important to note here that all timing stacks, including GPS based time-sync, are uniquely sensitive to \textit{delay} in the information transfer. To elaborate further, consider Figure~\ref{fig:delay-attack} that shows two agents $U$ and $T$, where the former requests time from the latter. Under non-malicious settings, $U$ has the same view of time as $T$ assuming instantaneous request and time transfer. However, an adversary that manages to intercept the time transfer could delay it by an arbitrary time $\delta t$. In this case, the time as seen by $U$ has an error of $\delta t$ with respect to the reference time maintained by $T$. \textit{Delays associated with time transfer affect the integrity of the information being transferred}. 

To conclude, we note that above case studies underscore the multifaceted nature of timing attacks that can be launched not just by a network based adversary (as discussed in section~\ref{sec:introduction}) but also by exploiting \textit{physical side channels} (\textbf{\texttt{C1,C5}}) and \textit{system software vulnerabilities} (\textbf{\texttt{C2-4}}).

\begin{table}
\scriptsize
\centering
\begin{tabular}{| p{2cm} | p{0.8cm} | p{0.9cm} | p{0.9cm} | p{0.85cm} | p{0.5cm} |}
 \cline{1-6}
\multicolumn{1}{|c|}{} & \multicolumn{1}{c|}{} & \multicolumn{4}{c|}{\textbf{Attacker Characteristics}}  \\
 \cline{3-6}
\textbf{\makecell[l]{Case Study}} & \textbf{\makecell[l]{Attack\\Type}} & \textbf{\makecell[l]{Access\\Type}} & \textbf{\makecell[l]{Control}} & \textbf{\makecell[l]{Stealth}} & \textbf{\makecell[l]{DoS}} \\
 \hline
 \makecell[l]{Industrial Sabotage} & \textbf{\texttt{\makecell[l]{A2}}} & \makecell[l]{Physical}  & \makecell[l]{\emptycirc} & \makecell[l]{\fullcirc} & \makecell[l]{\halfcirc} \\
 \hline
 \makecell[l]{Attack on\\AV-perception} & \textbf{\texttt{\makecell[l]{A2}}} & \makecell[l]{Privileged\\Execution}  & \makecell[l]{\emptycirc} & \makecell[l]{\halfcirc} & \makecell[l]{\fullcirc}  \\
 \hline
 \makecell[l]{Database\\Degradation} & \textbf{\texttt{\makecell[l]{A3}}} & \makecell[l]{Network\\Device} & \makecell[l]{\fullcirc} & \makecell[l]{\fullcirc} & \makecell[l]{\emptycirc}  \\
 \hline
\makecell[l]{Compromised Smart\\Contract Systems} & \textbf{\texttt{\makecell[l]{A1}}} & \makecell[l]{Physical/\\Priv. Exec.} & \makecell[l]{\fullcirc} & \makecell[l]{\fullcirc} & \makecell[l]{\emptycirc}  \\
 \hline
 \makecell[l]{GPS Manipulation} & \textbf{\texttt{\makecell[l]{A3}}} & \makecell[l]{Network/\\Physical} & \makecell[l]{\fullcirc} & \makecell[l]{\halfcirc} & \makecell[l]{\emptycirc}  \\
 \hline
\end{tabular}
\caption{Capabilities of adversaries from our case studies: full, half and empty circle indicates adversary has the particular capability, has it partially and no capability, respectively.}
\label{tab:case-study-attack-requirements}
\end{table}

\begin{table*}
\scriptsize
\centering
\begin{tabular}{| p{3.5cm} | p{2.75cm} | p{1.95cm} | p{2cm} | p{2.4cm} | p{1.55cm} |}
 \cline{1-6}
\multicolumn{1}{|c|}{} & \multicolumn{3}{c|}{\textbf{Timing Stacks}} & \multicolumn{1}{c|}{} & \multicolumn{1}{c|}{}  \\
 \cline{2-4}
\textbf{\makecell[l]{Applications}} & \textbf{\makecell[l]{Time Source}} & \textbf{\makecell[l]{Platform Software}} & \textbf{\makecell[l]{Sync. Protocols}} & \textbf{\makecell[l]{Comm. Tech./Medium}} & \textbf{\makecell[l]{Network Scale}} \\
 \hline
 \makecell[l]{GPS Satellites} & \makecell[l]{Atomic Clocks} & -  & \makecell[l]{Laser Ranging} & \makecell[l]{Microwaves} & \makecell[l]{Global} \\
 \hline
 \makecell[l]{Astronomical Telescopes} & \makecell[l]{Atomic Clocks} & -  & - & - & -  \\
 \hline
 \makecell[l]{Underwater Surveillance Networks} & \makecell[l]{Chip Scale Atomic Clocks} & - & - & \makecell[l]{Wired} & \makecell[l]{Continent}  \\
 \hline
\makecell[l]{Underwater Sensor Networks} & \makecell[l]{Quartz Crystals} & \makecell[l]{RTOS} & \makecell[l]{\{PCDE/MM\}-Sync} & \makecell[l]{Acoustic} & \makecell[l]{several Km}  \\
 \hline
\makecell[l]{Cellular Networks} & \makecell[l]{Quartz Crystals (TCXO)} & \makecell[l]{GPOS}  & \makecell[l]{PTP, NTP} & \makecell[l]{Ethernet} & \makecell[l]{City Scale}  \\
 \hline
 \makecell[l]{Data Centers} & \makecell[l]{Quartz  Crystals (TCXO)} & \makecell[l]{GPOS, Hypervisor}  & \makecell[l]{PTP,  NTP, Huygens} & \makecell[l]{Ethernet} & \makecell[l]{$~100$ meters}  \\
 \hline
  \makecell[l]{Smart Homes} & \makecell[l]{Quartz Crystals} & \makecell[l]{RTOS}  & \makecell[l]{FTSP, TPSN} & \makecell[l]{BLE, ZigBee} & \makecell[l]{$<100$m}  \\
 \hline
  \makecell[l]{Wireless Body Area Networks} & \makecell[l]{Quartz Crystals} & \makecell[l]{RTOS}  & \makecell[l]{NTP, FTSP} & \makecell[l]{BLE} & \makecell[l]{$\sim 1$m}  \\
 \hline
\end{tabular}
\caption{Time stack composition used by various CPS. They are composed of different technologies, however, they each have a time source, an operating system that manages this time source and a time synchronization protocol.} 
\label{tab:time-stack-diversity}
\end{table*}
\section{Systematization}\label{overview}

Conducting an exhaustive security analysis of the timing stack in modern CPS presents numerous challenges. The research literature has yet to fully assess the impact of adversarial manipulations on the timekeeping and measurement components of the timing stack (Figure~\ref{fig:time-stack-example}). This gap necessitates a comprehensive review of broader CPS security concerns and their connections to timing stack integrity. While existing studies often employ limited threat models focusing on specific timing security aspects, our approach leverages a broad threat model for a systematic examination. It requires us to evaluate security solutions designed for narrower threat models against a broader one. Additionally, threat models in current research may include caveats that complicate our analysis, such as Chronos~\cite{net-sync-chronos}, which proposes a secure multi-path NTP solution but overlooks vulnerabilities introduced by intermediary devices like home routers. Moreover, security enhancements for protocols can inadvertently introduce new vulnerabilities; for instance, Chronos' DNS attack mitigation strategy may inadvertently simplify these attacks~\cite{pitfalls-chronos}. This underscores the importance of meticulous analysis to ensure no legitimate threats are overlooked. Next, we describe the adversary model underpinning our study, outline our work's scope, and present the systematization of framework employed to cope with these challenges.

\noindent\textbf{Threat model.} We adopt a adversary model that comprehensively analyzes threats to the timing stack. We assume adversary's manipulation of the victim's perception of time to either be its primary objective or a means to undermine other system functionalities. The adversary may have following types of access to the victim: i) \textit{physical device}, ii) remote \textit{privileged code execution} and iii) \textit{control of a network device} on the path between the timing server and the client. Secondary goals of such adversary may include: i) launching a \textit{control}led attack e.g., controlling the extent of timing uncertainty introduced at the victim, ii) staying \textit{stealthy} in order to launch sustained attacks, or iii) rendering the timing service unusable (\textit{denial of service}--DoS attack). In Table~\ref{tab:case-study-attack-requirements}, we use our threat model to characterize adversaries from the above case studies (section~\ref{subsec:case-studies}).

\noindent\textbf{Analyzed Time Stacks.} Timing stacks exhibit significant diversity across application domains and with respect to underlying technologies (see table~\ref{tab:time-stack-diversity}). For high-precision requirements, GPS satellites employ atomic clocks as their \textit{time source}, whereas applications with relaxed timing precision, such as wireless sensor networks, might utilize quartz crystals. Similarly, \textit{local clock} on desktop computers, programmable logic controllers (PLCs) and cloud servers are maintained by a general purpose OS, a real-time OS and a hypervisor, respectively. And for distributed applications, time synchronization protocol vary with the network's size and topology. NTP~\cite{ntpv4-rfc} is the default time-sync protocol in wide area networks (WANs) using TCP/IP networking stack. Local Area Networks, such as data centers, may employ PTP~\cite{ptp-std-doc} for high accuracy synchronization, and resource constrained sensor networks may utilize RBS~\cite{Elson2003RBS}. Our work presents a unified framework for analyzing the security of these and other timing stacks, irrespective of the application domain or the adopted technologies.

\begin{figure}[tb]
    \small
    \centering
    \includegraphics[angle=270,width=0.5\columnwidth]{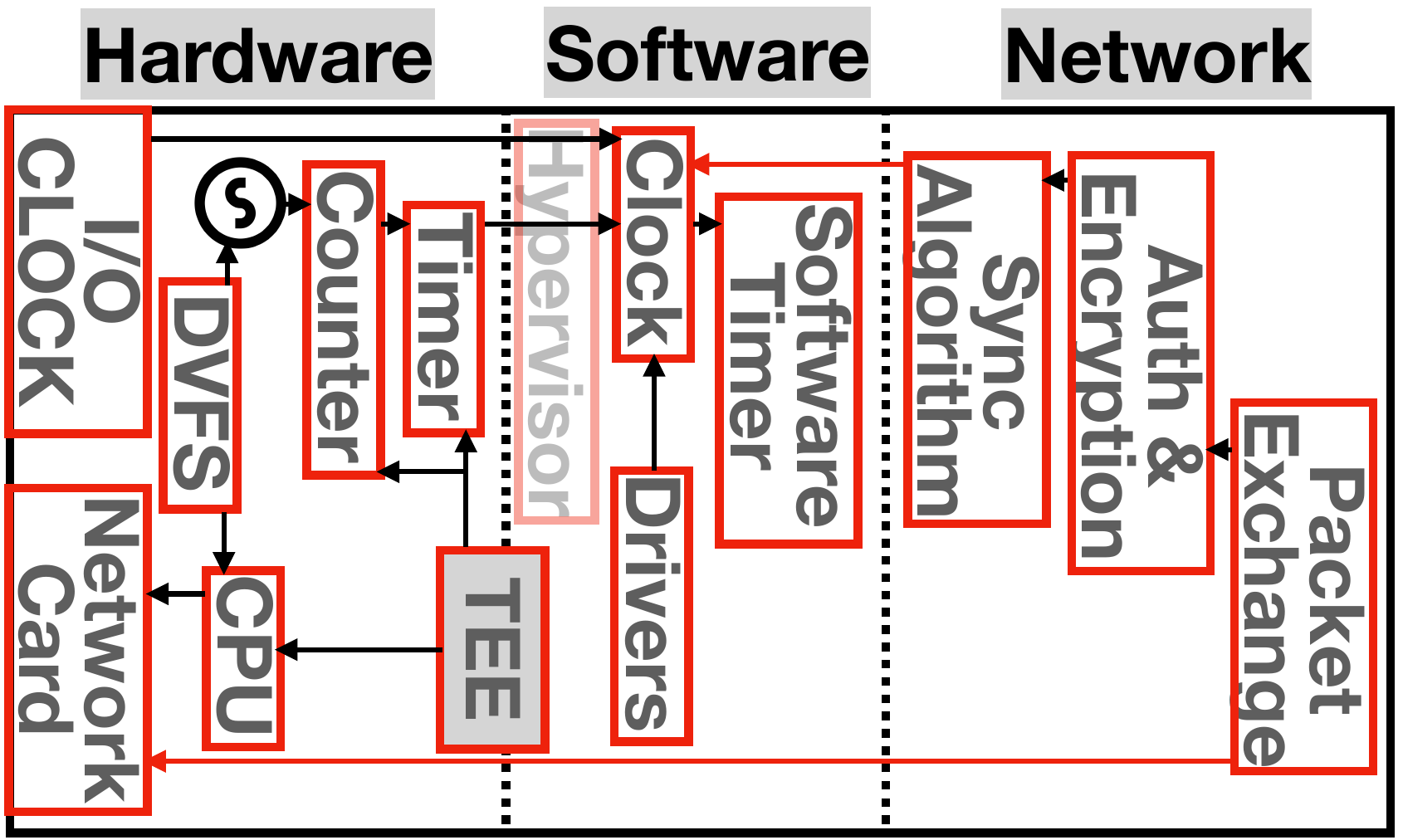}
    \caption{Various components of a typical CPS and their interactions that may be exploited by an adversary to attack its time stack (in red).}
    \label{fig:time-stack-attack-surface}
\end{figure}

\noindent\textbf{{Systematization Framework.}}
To analyze diverse timing stacks, we identify three layers common to all timing stacks, depicted in Figure~\ref{fig:time-stack-attack-surface}. 1) The \textbf{hardware layer} which contains the primary timing source for the system and the time measurement circuitry i.e. counter and timers. An adversary with \textit{physical access} to the target system may leverage physical side channels or hardware design limitations to manipulate this time measurement infrastructure. 2) The \textbf{software layer} maintains standardized clock and timer abstractions. Often these abstractions, implemented by the system software (OS, Hypervisor etc.), are vulnerable to attacks by \textit{adversaries that gain control of the system by exploiting software bugs}. 3) The \textbf{network layer} is responsible for aligning system time to an external reference. It is susceptible to manipulation by \textit{external adversaries having control of a device on the network path between the victim and the time server}.

We use this systematization framework to offer an exhaustive evaluation of security vulnerabilities within each layer of the timing stack in the forthcoming sections.
\section{Hardware Issues} \label{sec:hardware}
This section delves into vulnerabilities of the timing hardware, highlighting how such weaknesses can be exploited to alter a system's perception of time.

\subsection{Physical Side Channels} 
The physical components of the timing stack are susceptible to side channel attacks such as fault injection and manipulation of the system's thermal characteristics. To exploit these vulnerabilities, the adversary must have \textit{physical access} to the target system.

\noindent\textbf{\texttt{I01.} Laser Based Attacks on Crystal Oscillators.} Quartz crystal oscillators, integral to systems-on-chip (SoCs), are susceptible to optical laser attacks. Research by Kohei et al.~\cite{redshift} demonstrates a direct correlation between an oscillator's frequency and power of the laser incident upon it. 
They used this exploit to extract cryptographic keys embedded in the hardware. However, an adversary can use the same mechanism to alter the frequency of the oscillator that drives the hardware counters and timer registers on the SoC (\textit{time warping} attack --  \textbf{\texttt{A2}}). Fundamentally, this attack is possible as a result of crystal frequency sensitivity to ambient temperature (a laser incident on the crystal raises its temperature). Other time sources (e.g. atomic clocks) are also susceptible to environmental factors (temperature~\cite{nasa-atomic-clock}) and could be exploited by malicious adversaries albeit using a different attack mechanism than described here.

\noindent\textbf{\texttt{I02.} Computational Faults against Local Clock.} Computational faults resulting from under-volting a processor core affect both x86~\cite{hardware-plundervolt, hardware-V0ltpawn, hardware-voltage-pillager} and ARM~\cite{clock-sync-fault-presence, hardware-volt-jockey}; two of the most popular processor architectures. Such faults can corrupt instruction execution results (or even skip instructions altogether) and can be exploited to launch attacks against the \textit{local clocks} maintained by the system software. As described in section~\ref{subsec:local-clocks} (Time Keeping), system tick updates to the \textit{local clock} compute a new timestamp. Under-volting a processor core during this system tick update may introduce faults and random errors in the calculated timestamp (uncertainty in local time \textbf{\texttt{A3}}). For a successful attack, the adversary must predict when these computation are going to take place. An attacker using a physical interface for the attack~\cite{hardware-voltage-pillager}, can do so by monitoring the timer interrupt pin on the SoC. Finally, note that these computational faults can also be induced using other methods such as laser fault injection and voltage glitches using physical probes~\cite{hardware-plundervolt, hardware-V0ltpawn}~\footnote{Research has demonstrated that the under-volting attacks are also possible through a software interface only~\cite{hardware-voltage-pillager} and a remote adversary with \textit{privileged access} can also launch this attack.}.

\subsection{Design Limitations}
The trade-off between security and performance often de-prioritizes the former in system design, potentially exposing timing mechanisms to exploitation due to design oversights or flawed assumptions. Exploiting these design flaws often require the adversary to have \textit{privileged access} to the system software.

\noindent\textbf{\texttt{I03.} Exploiting Energy Management Mechanisms.} The prevalent Dynamic Voltage and Frequency Scaling (DVFS) mechanisms in CPS for energy efficiency inadvertently introduces a vulnerability. Typically, a dedicated system software module oversees energy management and controls the DVFS interface. However, any software with escalated privileges can access this interface. This also includes an attacker who gained escalated privileges by exploiting system software bugs (table~\ref{tab:cve-stats}). Such an adversary uses the DVFS interface to alter the system frequency without alerting the \textit{local clock}. Being unaware of the change, system software relies on an outdated clock frequency to convert time from clock cycles to wall clock time~\footnote{As detailed in section~\ref{subsec:local-clocks}, system software uses clock frequency value to translate hardware time measurements from cycle count to seconds}. This attack introduces \textit{time warping} (\textbf{\texttt{A2}}) to the \textit{local clock} and the extent of this warping can be \textit{precisely controlled} by the \textit{stealthy} adversary.

\noindent\textbf{\texttt{I04.} Re-configurable timing counters.} 
System software uses hardware counters such as Intel's TSC~\cite{intel-tsc} and ARM's CNTVCT~\cite{arm-cntvct} for updating the \textit{local clock} (section~\ref{subsec:local-clocks}--Time Keeping). In modern systems, these counters are write-protected and does not allow the operating system to manipulate them directly. However, with the implementation of virtualization extensions, both architectures introduce an offset register~\cite{arm-cntvoff, intel-tsc-offset} shown in figure~\ref{fig:software-stime-stacks}. This register, originally designed to allow virtualization software to emulate counters for multiple guests, is writable. A malicious agent with \textit{privileged execution} access can exploit this offset register to manipulate the system's time view and induce \textit{time travel} (\textbf{\texttt{A1}}) in the local clock. Older systems such Intel processors designed before $2011$~\cite{intel-variant} provide writable counters and are even more susceptible to adversarial attacks as it allows the adversary to launch without having to rely on virtualization extensions which may or may not be enabled by default.
\section{Software Issues} 
This section examines the security vulnerabilities of the software layer of CPS, that could be exploited to attack their timing stack. These vulnerabilities are exploited by an adversary with \textit{privileged access} to the system. 

\begin{table}
\scriptsize
\centering
\begin{tabular}{
 | c | c | c | c | }
 \hline
  \textbf{System} & \textbf{Application Domain} & \textbf{Total} & \textbf{Critical} \\
 \hline
 \hline
Xen  & Virtualization & $173$ & $74$   \\
 \hline
 \hline
 Linux Kernel & Virtualization, Desktop, Embedded & $1198$ & $549$   \\
 \hline
 Android  & Embedded & $4574$ & $1971$  \\
 \hline
 iOS & Embedded & $1557$ & $969$   \\
 \hline
 FreeRTOS  & Low End Embedded & $14$ & $4$   \\
 \hline
 \hline
  Nvidia Tegra  & TEE & $22$ & $17$   \\
 \hline
  Linaro OPTEE  & TEE & $50$ & $7$   \\
 \hline
\end{tabular}
\caption{List of vulnerabilities discovered in system software used by different application domains~\cite{cve-details}.}
\label{tab:cve-stats}
\end{table}

\subsection{System Software Bugs} The vulnerabilities of the system software often lead to a non-privileged adversary to gain escalated privileges~\cite{sanitizing-for-security}. With escalated privileges, the adversary also gains ability to attack the \textit{local clock}.

\noindent\textbf{\texttt{I05.} Privilege Escalation Attacks.} System software provides interfaces to the time-sync protocol (network layer) to update the \textit{local clock}~\cite{linux-adjtime} and align it with the network time. While this capability is crucial for time synchronization, it can also be exploited by an adversary with elevated privileges to manipulate system's time. Such an attacker is free to launch \textit{time travel} (\textbf{\texttt{A1}}), \textit{time warping} (\textbf{\texttt{A2}}) or \textit{random error} (\textbf{\texttt{A3}}) attacks using this interface. This attack is fundamentally enabled by the privilege escalation vulnerabilities that are prevalent in commodity systems. Table~\ref{tab:cve-stats} provides a list of vulnerabilities discovered in system software of various platforms between 2018 and 2023, each representing a potential threat to the system's timing stack. Note that the attacker does not need to find a new vulnerability in the victim's software. It can also exploit privilege escalation vulnerabilities discovered by others but not yet patched by the system administrator.

\begin{figure}[h]
    \centering
    \includegraphics[scale=0.12]{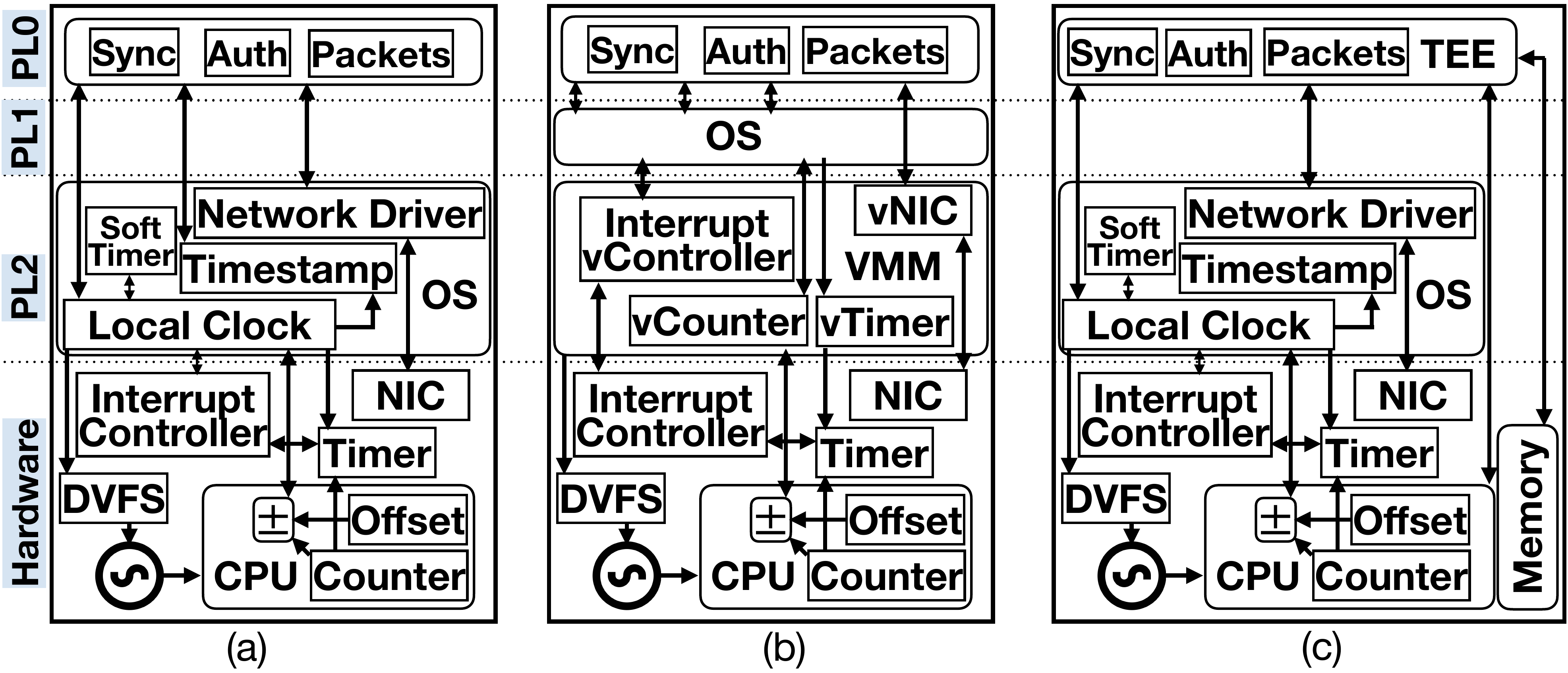}
    \caption{Attack surfaces: a) general purpose platforms, b) virtualized environments and c) with a unprivileged TEE.}
    \label{fig:software-stime-stacks}
\end{figure}

\noindent\textbf{\texttt{I06.} Untrusted device drivers.} On most systems, device drivers execute with privileged access, sharing the same context as the operating system (figure~\ref{fig:software-stime-stacks}). Drivers installed from untrusted sources may contain malicious code~\cite{sok-attacks-on-software-supply-chains} that executes with elevated privileges and may launch attacks against the timing stack. It can do so via one of the following mechanisms: i) alter the \textit{timing counters} ($I04$) used by the \textit{local clock}. When timing counters are protected, ii) it would locate the system clock data by scanning physical memory (leveraging its privileged position). Once located, it can manipulate the \textit{local clock}. If the physical memory scan is infeasible, iii) the malicious driver may register an interrupt and configure it to trigger frequently. It seeks to intercept and delay system tick updates for the \textit{local clock}. The first two of these mechanisms allow the adversary \textit{precise control} to launch any of the attacks (\textbf{\texttt{A1-3}}) described in section~\ref{subsec:attack-types}. While, the last mechanism offers \textit{less control} and is likely to result in \textit{increased timing error} (\textbf{\texttt{A3}}).

\noindent\textbf{\texttt{I07.} Virtual timer interrupts in the cloud.} Hypervisor is the most privileged software in virtualized environments and enables hardware resource sharing among multiple operating systems (guests). As shown in the figure~\ref{fig:software-stime-stacks}(b), it provides virtual instances of the physical hardware components to each guest. These emulated components also include hardware counters and timer interrupts~\cite{time-stack-hyp-paratick}. Hypervisor, just like an OS, contains bugs (see Xen in table~\ref{tab:cve-stats}) and may get compromised by a malicious agent. Such a hypervisor may manipulate the emulated hardware to alter guests' view of time. For instance, it can launch \textit{time warping attack} (\textbf{\texttt{A2}}) by delaying the timer interrupt and slow down the counter, or vice versa, without notifying a guest. Beyond \textit{warping guests' time}, the hypervisor can also manipulate time by shifting it backward or forward (see $I04$). While the specific mechanisms may vary across platforms, the inherent design of virtualized platforms enables a malicious hypervisor to launch timing attacks.

\subsection{TEE Limitations}\label{subsec:tee-limitation}
\begin{figure}[t]
    \small
    \centering
    \begin{subfigure}[t]{0.225\columnwidth}
        \centering
        \includegraphics[width=\columnwidth]{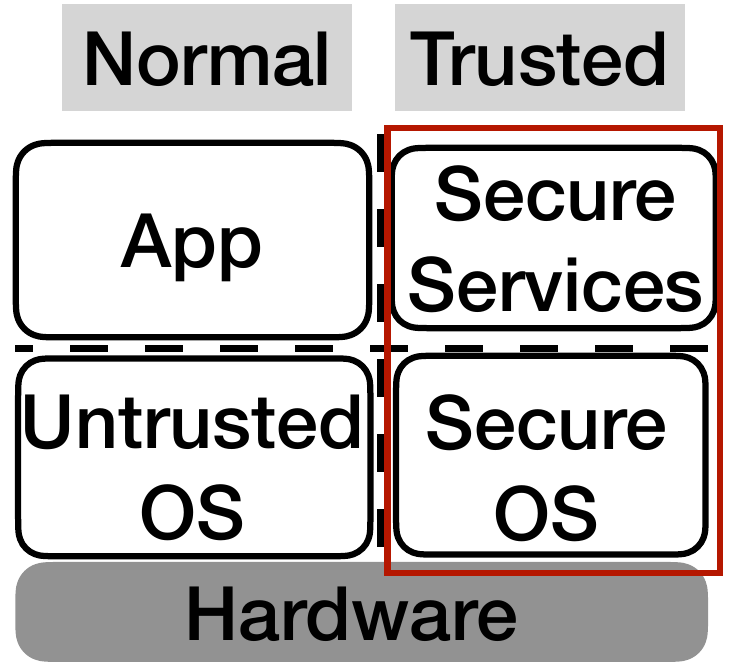}
        \caption{}
        \label{fig:privileged-tee}
    \end{subfigure}
    \hspace{20pt}
    \begin{subfigure}[t]{0.215\columnwidth}
        \centering
        \includegraphics[width=\columnwidth]{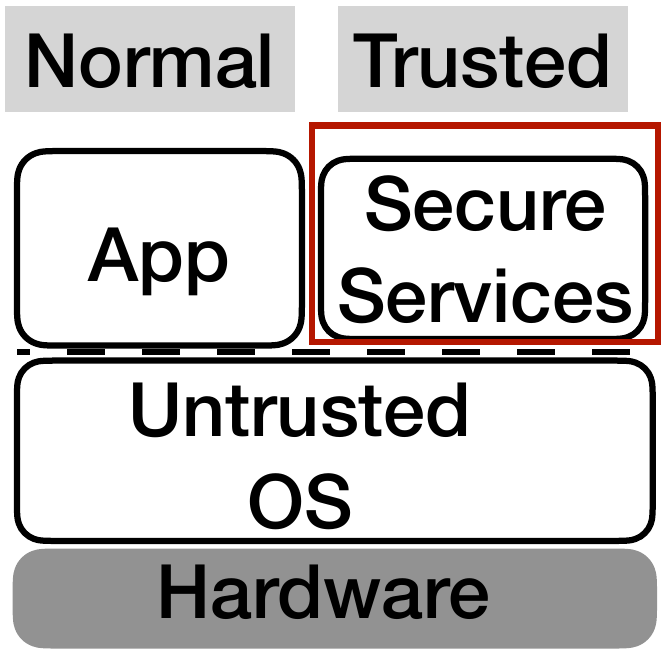}
        \caption{}
        \label{fig:userspace-tee}
    \end{subfigure}
    \caption{a) Privileged TEE design; trusted software has direct (secure) access to hardware resources. b) Un-privileged TEE design; cannot securely access any hardware resource except memory.}
    \label{fig:tee-paradigms}
\end{figure}

Hardware-based TEEs are designed under a strong threat model, treating the system software (e.g., OS, hypervisor etc.) as potentially malicious. These TEE designs follow two paradigms: i) user-space TEEs, exemplified by Intel SGX~\cite{intel-sgx-explained}, and ii) privileged TEEs, like ARM Trustzone~\cite{sok-trustzone-cves} (see Figure~\ref{fig:tee-paradigms}). Both designs, however, aim to secure sensitive code and data. Enabling a secure time stack within these enclaves have been a challenge; especially in the case of the former due to the limited hardware access~\cite{time-stack-abouttime}. 

\noindent\textbf{\texttt{I08.} TEE Design Limitations.}
User-space TEEs (HETEE, Fidelious, HIX, Intel SGX) protect the user application's code and data via cryptography~\cite{sok-hardware-tee}. However, they fall short of securing the time stack due to limited hardware access. The TEE software's access to system's hardware resources such as interrupts, timers and network devices is mediated by the untrusted OS (see figure~\ref{fig:software-stime-stacks}c). A compromised OS can intercept and manipulate the TEE software's access to the timing resources. For instance, timing API $sgx\_get\_trusted\_time$ provided by Intel SGX, a user-space TEE, is vulnerable to delay attacks by a compromised OS~\cite{time-stack-timeseal}. These attacks are launched by directly exploiting lack of direct TEE access to the hardware counter causing increased uncertainty in SGX time (\textbf{\texttt{A3}}). Newer SGX iterations mitigate this by enabling direct $TSC$ (hardware counter) access by the TEE software. However, this still does not enable a secure time stack as the $TSC$ is not fully secure and is vulnerable to the compromised OS (see $I04$). Beyond local time, the lack of TEE direct access to network card also prevents it from obtaining trusted time through the network. The TEE's network traffic is handled by an untrusted network driver (see figure~\ref{fig:software-stime-stacks}c) which can add arbitrarily delay timing packets and induce uncertainty in the timing information (\textbf{\texttt{A3}}) received by the TEE.

\noindent\textbf{\texttt{I09.} Compromised TEE Software.}
Privileged TEEs such as ARM Trustzone does have access to a secure counter and timer, allowing the TEE software to maintain secure \textit{local clock}. Unfortunately, this secure clock may still get exposed to adversaries because of the TEE software vulnerabilities which are discovered regularly~\cite{sok-sgx-fail, sok-trustzone-cves} (see Nvidia Tegra \& Linaroo OPTEE in table~\ref{tab:cve-stats}). Many TEE vulnerabilities allow adversary to manipulate code inside the TEE~\cite{boomerang-trustzone} and put the adversary in control of the previously isolated resources of the TEE including its \textit{local clock}. Such an attacker can \textit{hide} itself with relative ease and launch any of the attacks \textbf{\textit{A1-3}} discussed in the section~\ref{subsec:attack-types}. It is important to note that if the TEE software is compromised, an application can no longer trust any timing stack on the system as it already does not trust timing stack maintained by the untrusted OS. 

\section{Network Issues} \label{sec:network-issues}
This section delves into the network layer's vulnerabilities, pivotal for synchronizing time across digital systems. Such synchronization is vital for applications ranging from digital payments to industrial automation. Yet, it faces threats from \textit{attackers controlling network devices} (on-path attacker) or \textit{possessing privileged access to a victim's local network stack} (off-path attacker).

\subsection{Limited Use of Authentication Mechanisms} Cryptography techniques, used by protocols like NTP~\cite{ntpv4-rfc} and PTP~\cite{ptp-std-doc}, play a critical role in ensuring data integrity and origin authentication of the time-sync traffic, thwarting man-in-the-middle (MITM) attacks. Yet, several issues persist regarding the adoption of these methods making time-sync protocols vulnerable to attacks.

\noindent\textbf{\texttt{I10.} False packet injection.} A MITM adversary can impersonate a genuine time server and send false time-sync packets to the target. These attacks may result from weak assumptions underlying the authentication mechanism adopted by the time-sync protocol. For instance, the reliance of NTP's broadcast mode authentication protocol TESLA~\cite{tesla-cryptography} (also used by PTP~\cite{ptp-std-doc}) on loosely synchronized devices creates a circular dependency between authentication and time-sync~\cite{ntp-replay-drop-attack}, rendering the former useless. Moreover, infiltration of malicious servers in the pool of legitimate time servers is  a genuine concern~\cite{shark-ntp-pool, devil-time-origin}. It is because cryptographic authentication only protects against a MITM attacker and the malicious servers render it ineffective. This allows Kwon et. al., to use a handful of malicious time servers, injected to the NTP pool~\cite{ntpd-pool-project}, to disrupt time-sync clients spread over entire countries~\cite{shark-ntp-pool}. Despite their shortcomings, authentication techniques make packet injection attacks harder. However, the adoption of these mechanisms is not universal. For instance, Huygens~\cite{huygens}, RBS~\cite{Elson2003RBS}, FTSP~\cite{ftsp-2004}, TPSN~\cite{tpsn-2003} do not implement any origin authentication mechanisms and have no protection against packet injection. The severity of the issue is evident from the fact that \textit{RBS, FTSP and TPSN} are among the most cited protocols for time-sync in sensor networks. In contrast, secure time synchronization protocols such as the one introduced by Ganeriwal et. al.~\cite{net-sync-wsn-sec-prot} has received an order of magnitude fewer citations (see table~\ref{tab:time-sync-wsn-citations}). Packet injection is one of the most potent attacks against time-sync protocols and could be used to induce \textit{time travel, warping or just increased uncertainty} (\textbf{\texttt{A1-3}}) in the victim's view of time.

\begin{table}
\scriptsize
\centering
\begin{tabular}{ | c | c | c | c | }
 \hline
  \textbf{Protocol} & \textbf{Authentication} & \textbf{Date Published} & \textbf{Citations} \\
 \hline
 \hline
  RBS~\cite{Elson2003RBS}  & \textit{No} & $2003$ & $3927$   \\
 \hline
   TPSN~\cite{tpsn-2003}  & \textit{No} & $2003$ & $3206$   \\
 \hline
   FTSP~\cite{ftsp-2004}  & \textit{No} & $2004$ & $3052$   \\
 \hline
 \hline
   Secure Time-Sync~\cite{Elson2003RBS}  & \textit{Yes} & $2005$ & $278$   \\
 \hline
\end{tabular}
\caption{One of the earliest time-sync protocols proposed for wireless sensor networks (WSNs). The protocols (RBS, TPSN and FTSP) that do not incorporate authentication mechanisms have received an order of magnitude more citations than the protocol (STS) that make use of cryptography mechanisms. \textit{Source: Google Scholars as of Jan 22, 2024.}}
\label{tab:time-sync-wsn-citations}
\end{table}

\noindent\textbf{\texttt{I11.} Packet modification.} Correct implementation of authentication protocols prevents false packet injection but may not prevent against packet modification. This is best exemplified by PTP, which makes use of authentication~\cite{ptp-std-doc} to protects the PTP packets except the correction field of the packet header. This field allows each network node to update correction field with the packet processing delay. PTP uses this information to achieve better time-sync accuracy by eliminating the variable network delays~\cite{net-sync-ptp-covert-channel}. However, a MITM attacker (on-path or off-path) can add incorrect information to this field and manipulate the PTP client. Jacobs et, al., use this channel to introduce \textit{significant offsets} (\textbf{\texttt{A3}}) to the victim device while \textit{avoiding detection}. They could also induce the victim device to change its clock frequency (\textbf{\textit{A2}}), resulting in an even larger time deviation from the time server~\cite{net-sync-ptp-covert-channel}. We note that this attack is not PTP specific, and any time-sync protocol seeking network delay information may be subject to this attack. Finally, we also note that this technique is less sophisticated as it does not require by-passing authentication requires.

\noindent\textbf{\texttt{I12.} Packet replay.} Authentication issues discussed in $I10$ can also result in replay attacks. In this attack, the adversary repeatedly sends one or a sequence of pre-recorded time server packets to the victim. Packet replay attacks have been successfully demonstrated against NTP broadcast mode~\cite{ntp-replay-drop-attack}. Malhotra et. al. exploited limitations in existing NTP client implementations to keep the victim stuck at a single point in time (\textbf{\texttt{A1}}). They point out that the one-way nature of the time-sync traffic (NTP broadcast mode) enables this attack. It implies that other one-way time synchronization protocols such as RBS~\cite{Elson2003RBS} may also be susceptible to this attack.

\noindent\textbf{\texttt{I13.} Spoofing Wireless Timing Signals.} Time-sync protocols like GPS, ROCS~\cite{ROCS-FM-Beacons}, Syntonizor~\cite{Syntonizor-AC-powerlines} and WizSync~\cite{WizSync-Wifi-Beacons} work using a periodic wireless timing signal that is transmitted directly from the timing source(s) to the clients i.e. over a single hop. These protocols lack authentication mechanisms allowing adversaries to spoof timing signals. This attack is the equivalent to packet manipulation attack on packet exchange based protocols (NTP~\cite{nts-rfc}, PTP~\cite{ptp-std-doc}, FTSP~\cite{ftsp-2004} etc.). Similar to the packet manipulation attacks, an external adversary mimics a trusted timing source but transmits incorrect timing information. It does so by generating a powerful spoof signal, using antenna(s), that can overpower the legitimate signal. Such an attacker often stays \textit{stealthy} while introducing uncertainty in the victim's local clocks~\cite{gps-spoofing-fundamentals} (\textbf{\texttt{A3}}). Satellite based global positing systems (GPS) is a typical target of this attack~\cite{gps-spoofing-21}. However, other time-sync protocols in this category (e.g., ROCS,~\cite{ROCS-FM-Beacons}, WizSync~\cite{WizSync-Wifi-Beacons} and Syntonizor~\cite{Syntonizor-AC-powerlines} etc.) haven't seen significant spoofing attacks due to their limited application. Nevertheless, signal spoofing remains a viable attack option for a motivated adversary.

\subsection{Availability Issues}
Beyond modifying timing packets, time-sync is also affected by just delaying the transmission of the timing information (as discussed in section~\ref{subsec:case-studies}). An adversary may leverage this observation and use unpredictable delays to add errors to the time-sync process or it  may outright block time-sync traffic headed towards the victim. 

\noindent\textbf{\texttt{I14.} Packet delay.} Time synchronization protocols determine the time offset between the server and the client by exchanging packets over the network. These network packets experience delays causing uncertainty in the exchanged timing information and the corresponding offset calculations (see section~\ref{subsec:case-studies}). Time-sync protocols rely heavily on precise network delay measurements to remove this uncertainty in the offset estimations. NTP~\cite{ntpv4-rfc} solves this challenge by measuring round trip times (RTTs) and computes network delay as half of the RTT, assuming symmetric delays~\cite{rfc1305}. On the other hand PTP measures the network delays by mandating each processing node to update the PTP packets with its resident delay (see $I11$). While effective under normal network conditions, these delay estimation mechanisms are not robust to adversarial delays. A malicious network node may introduce additional network delays~\footnote{In case of NTP, the server-bound and client-bound packets are delayed by different duration while for PTP the adversary would not update PTP packets with its resident delay} to degrade the synchronization performance. For instance, Annesi et. al. show that delay attacks against PTP can induce errors of several milliseconds, accumulating over time to even larger values under a sustained attacks~\cite{ptp-futile-encryption} (\textbf{\texttt{A2}}). However, vulnerability to delay attacks extend beyond NTP and PTP; virtually all time-sync protocols are susceptible to these attacks.

\noindent\textbf{\texttt{I15.} Packet drop.} 
Intercepting and dropping time-sync packets is a simple yet effective MiTM attack that desynchronizes the victim device from its time server. Facing this \textit{denial-of-service} attack, the victim solely relies on its \textit{local clock} which diverges away from the server time (\textbf{\texttt{A3}}) dictated by the stability of the victim's time source. For low-end systems using inexpensive quartz crystals, the time difference may accumulate to several minutes per day. In contrast, devices using more stable oven-controlled quartz oscillators may experience deviations of only a few seconds in the same period. However, despite its effectiveness, the victim can deduce potential instances of this attack, with relative ease, from sudden unavailability of time-server.

\noindent\textbf{\texttt{I16.} Blocking Wireless Timing Signals.} For single-hop wireless time synchronization (GPS~\cite{gps-spoofing-fundamentals}, ROCS~\cite{ROCS-FM-Beacons}, WizSync~\cite{WizSync-Wifi-Beacons} etc.), denial of service attack takes the form of blocking the wireless timing signal. An adversary achieves this by generating high powered noise in the frequency band used by the wireless timing signal. It requires physical proximity to the target and signal transmission equipment, raising the cost of this attack. Nevertheless, GPS signal blocking techniques have been studied extensively~\cite{gps-jamming-overview} due to ubiquitous use of GPS by defense and civil infrastructure. In principle, other single-hop wireless protocols such as Syntonizor~\cite{Syntonizor-AC-powerlines} and ROC~\cite{ROCS-FM-Beacons} are also vulnerable to these attacks, even though no such attack against them is known.

\subsection{Implementation Issues} In addition to the the communication medium, the end-points of this channel i.e. the applications implementing the time-sync protocol themselves represent an attack surface.

\noindent\textbf{\texttt{I17.} Untrusted time synchronization software.} Applications implementing time-sync protocols may harbor security vulnerabilities of their own. For instance, CVE database lists 98 vulnerabilities, discovered over the years, in the NTP application developed by \textit{NTP.org}~\cite{ntp-cve-details}. This application is used by both the time-sync clients and servers,~\footnote{It is recommended for servers joining the NTP pool project~\cite{ntpd-pool-project}.} and can be exploited by an adversary with access to \textit{privileged execution} on the victim device or \textit{a network connection to the NTP application}. An attack exploiting client side application vulnerability would only affect a single machine, however, the server side exploit would affect time alignment at all of its clients. Further, these attacks may cause the target applications to crash pausing the time-sync service or may just degrade time-sync performance (\textbf{\texttt{A3}}) over longer periods. It is worth pointing out time-sync applications executing in the privileged context present an even bigger risk, as any vulnerability in them could compromise the system beyond time-sync service.
\section{Defenses Against Timing Attacks}
This section examines defense mechanisms aimed at mitigating the vulnerabilities faced by the timing stack's three layers. Table~\ref{tab:defense-examples} provides representative examples of this work and the extent to which it addresses timing stack issues. In this analysis, we focus exclusively on system-based solutions for safeguarding timing architectures, deferring the discussion of theory-based approaches for the next section.

\begin{table*}[t]
\footnotesize
\centering
\begin{tabular}{ p{4.75cm}  p{1cm}  p{1.25cm}  p{1.25cm}  p{1.25cm}  p{1.5cm}  p{1.5cm}  p{1.5cm}  }
 \multicolumn{1}{c}{} & \multicolumn{2}{c}{Hardware Issues} & \multicolumn{2}{c}{Software Issues} & \multicolumn{3}{c}{Network Issues}\\
 \cmidrule(lr){2-3} \cmidrule(lr){4-5} \cmidrule(lr){6-8}
    & \textit{Side Ch.} & \textit{Flaw. Des.} & \textit{Soft. Bugs} & \textit{TEE Issues} & \textit{Auth. Issues} & \textit{Avail. Issues} & \textit{Impl. Issues} \\
 \hline
 Wei et. al.~\cite{lfi-ro-based} & \halfcirc & \emptycirc & \emptycirc & \emptycirc & \emptycirc & \emptycirc & \emptycirc \\
 Arm Generic Timer~\cite{arm-generic-timer} & \emptycirc & \halfcirc & \emptycirc & \emptycirc & \emptycirc & \emptycirc & \emptycirc \\
 TPM Counters~\cite{ftpm} & \emptycirc & \fullcirc & \emptycirc & \emptycirc & \emptycirc & \emptycirc & \emptycirc \\
 \hline
 Timeseal~\cite{time-stack-timeseal} & \emptycirc & \emptycirc & \halfcirc & \halfcirc & \emptycirc & \emptycirc & \emptycirc \\
 T3E~\cite{trusted-time-t3e} & \emptycirc & \emptycirc & \halfcirc & \halfcirc & \emptycirc & \emptycirc & \emptycirc \\
Scone~\cite{sandbox-scone} & \emptycirc & \emptycirc & \emptycirc & \halfcirc & \emptycirc & \emptycirc & \emptycirc \\
SeCloak~\cite{sandbox-secloak} & \emptycirc & \emptycirc & \emptycirc & \halfcirc & \emptycirc & \emptycirc & \emptycirc \\
\hline
Cryptographic Communications~\cite{net-sync-ptp-sec, ntpv4-rfc} & \emptycirc & \emptycirc & \emptycirc & \emptycirc & \halfcirc & \emptycirc & \emptycirc \\
Chronos~\cite{net-sync-chronos} & \emptycirc & \emptycirc & \emptycirc & \emptycirc & \halfcirc & \fullcirc & \emptycirc \\
Semperfi et. al.~\cite{gps-anti-spoofing-semperfi} & \emptycirc & \emptycirc & \emptycirc & \emptycirc & \halfcirc & \fullcirc & \emptycirc \\
\hline
\end{tabular}
\caption{Examples of representative papers that propose mitigation for timing stack issues (\textit{Ixx}). 
A full circle indicates the research paper mitigates all issues in the category, a half circle indicates that some of the issues in a category are addressed and an empty circle signifies the lack of proposed mitigation for the given category. Note that implementation issues ($I17$) are not addressed by system-based approach as they arise from errors in execution of this approach itself.}
\label{tab:defense-examples}
\end{table*}

\subsection{Securing the Hardware}
We begin by highlighting design solutions that address issues arising from the physical side channel and the design limitations of the timing circuitry.

\noindent\textbf{\texttt{D01.} Laser Fault Injection Countermeasures.} In response to the rising threat of laser fault injection (LFI), a significant body of research has focused on detection techniques~\cite{lfi-multi-spot, lfi-tdc, lfi-ro-based}. These efforts concentrate on identifying lasers incident on the SoC and provide countermeasures against laser-based computational faults ($I02$). However, these methods may not detect laser-based attacks on crystal oscillators which are external to the SoC. Nevertheless, they offer a promising starting point for developing countermeasures against laser attacks on crystal oscillators ($I01$). For instance, He et al. introduced a ring oscillator-based watchdog that analyzes clock signal irregularities to detect LFI attacks~\cite{lfi-ro-based}. Such solutions can be further developed to detect attacks on the external oscillators by analyzing the analog clock signal generated by them.

\noindent\textbf{\texttt{D02.} Monotonic and Fixed Frequency Counters.} Many contemporary System-on-Chips (SoCs) incorporate specialized counters that are monotonic i.e. consistently counting in a single direction and immune to resets. Further, the frequency of these counters remains independent of the DVFS mechanism, thwarting any attempt by a privileged adversary to exploit energy management interfaces for attacks on the timing stack ($I03$). ARM's generic timer~\cite{arm-generic-timer} and Intel's TSC are notable examples~\footnote{when virtualization extensions are disabled.}~\cite{intel-tsc}. Monotonic counters, uninfluenced by DVFS, are also prevalent in Trusted Platform Modules (TPMs) embedded in modern SoCs\cite{ftpm}. However, TPMs, situated outside processor chips, suffer from substantial access latency, constraining their utility~\cite{time-stack-timeseal}. These fixed rate monotonic counters present a substantial advance towards fixing hardware design issues that could enable timing attacks. However, despite these advancements, the challenge of counter manipulation still persists on legacy hardware and modern systems that incorporate virtualization extensions (see $I04$). 

\subsection{Software Defenses}
This subsection delves into solutions designed to protect the integrity of \textit{local clock}'s data. 

\noindent\textbf{\texttt{D03.} Trusted Timing Services.} Development efforts have been concentrated on integrating trusted timing services within TEEs to counter vulnerabilities in system software and device drivers ($I05, I06$). For instance, ARM's TrustZone provides a privileged TEE with direct access to a secure counter and timer~\cite{arm-generic-timer}, enabling TEE software to maintain a \textit{trusted local clock}. Intel SGX, a user-space TEE, benefits from solutions like Timeseal~\cite{time-stack-timeseal}, which secures the enclave's trusted timing API against delay attacks, and T3E~\cite{trusted-time-t3e}, utilizing secure TPM counters to provide trusted timing within the SGX enclave. These approaches for constructing a trusted timing stack are not confined to Intel and ARM-based TEEs but are extensible to other TEE architectures. Despite their significance, these trusted timing solutions exhibit several notable limitations: (i) the application code requiring trusted time must execute within the TEE's isolated environments, which enlarges the Trusted Computing Base (TCB), thereby increasing the security risk to the TEE software\footnote{TCB refers to the code and data that reside inside a TEE.}; and (ii) crucially, these trusted timing solutions do not incorporate secure time-synchronization services.

\noindent\textbf{\texttt{D04.} Trusted I/O.} Researchers have proposed several solutions to enable user-space TEEs' (Figure~\ref{fig:userspace-tee}) direct access to I/O devices ($I08$). One such effort, Aurora~\cite{sandbox-aurora}, allows Intel SGX to access high-resolution counters in the hardware, among other peripherals. SGX enclave's access to these counters can improve trusted time services such as Timeseal~\cite{time-stack-timeseal} when integrated with it. Other initiatives like Scone~\cite{sandbox-scone} and SGXIO~\cite{sgxio} provide secure network I/O to the SGX enclave, essential for synchronizing trusted time stacks inside TEEs with network time. However, they only safeguard the integrity and confidentiality of network packets, falling short of preventing delay attacks by untrusted system software.

\subsection{Secure Time Synchronization}
Time synchronization security has garnered significant focus within timing security research. Efforts in this domain have concentrated on enhancing various aspects of time-sync protocols to fortify them against adversarial actions.

\noindent\textbf{\texttt{D05.} Cryptographic Communications.} Time-sync protocols can enhance their security against packet manipulation attacks ($I10$, $I11$) by utilizing authentication and encryption, provided they adhere to the following conditions: (i) cryptographic functions should fully protect network packets~\cite{net-sync-ptp-covert-channel}, and (ii) their operation should not require time synchronization as a prerequisite~\cite{ntp-replay-drop-attack}. For example, the authentication and encryption mechanisms introduced by the NTS standard (RFC8915~\cite{nts-rfc}) for NTP's client-server mode fulfill both criteria. Conversely, PTP's authentication mechanisms fail to meet these conditions because: (i) the correction field in the PTP packet header remains unprotected by authentication, and (ii) the recommended authentication protocol, TESLA, requires devices to be loosely synchronized. Despite these shortcomings, such security measures constitute the primary defense against network adversaries and should be implemented by time-sync protocols to either partially or completely mitigate packet manipulation attacks. It is important to note that packet manipulation by malicious time servers~\cite{shark-ntp-pool} remains feasible and necessitates further defense mechanisms.

\noindent\textbf{\texttt{D06.} Multipath Time Transfer.} Delay attacks $I14$ cannot be mitigated using cryptographic mechanisms and must be prevented using other techniques. Using multiple paths for synchronization between two systems is one strategy to mitigate these attacks. This approach forces the attacker to identify and introduce delay along all the time-sync paths which is a significantly more challenging task. Mizrahi et al. propose a game-theoretic model for such a multipath time synchronization scheme~\cite{multi-path-game-theory}. Similarly, Chronos~\cite{net-sync-chronos} employs multi-path synchronization by querying reference time from various NTP servers. This approach is not only effective in mitigating delay attacks $I14$ but it also mitigates packet drop $I15$ attacks. However, the multi-path approach is ineffective against a malicious device that functions as bottleneck on the path between the client and the time server(s).

\noindent\textbf{\texttt{D07.} Algorithmic Updates.} Recent research has focused on algorithmic approaches for mitigating delay ($I14$) and replay ($I12$) attacks. For instance, Fatima et al.\cite{net-sync-feedforward} present a feedforward clock model, for PTP, along with an algorithm proficient in detecting delay-free packets. They remove rest of the packets that were potentially delayed or replayed by an adversary before estimating time-sync parameters (offset and skew). Likewise, Chronos\cite{net-sync-chronos} reinforces the multi-path time-sync approach using a byzantine fault tolerance-based algorithm for the random selection of time servers in each synchronization round. Adopting these algorithmic updates will enhance time-sync protocol's resiliency against delay attacks, but it does not completely prevent performance degradation~\cite{net-sync-feedforward}.

\noindent\textbf{\texttt{D08.} GPS Anti-jamming \& spoofing.} The prevalence of GPS jamming ($I16$) and spoofing ($I13$) attacks against critical navigation systems have motivated a large number of studies. These works have demonstrated the ability to acquire weak GPS signals amidst jamming, allowing for an average positioning error of 16m~\cite{gps-anti-jamming-post-correlation, gps-anti-jamming-post-wavelet}. Addressing GPS spoofing, Khalajmehrabadi et al. present detection techniques that estimate the extent of the spoofing signal, empowering devices to take necessary mitigation measures~\cite{gps-anti-spoofing-tsarm}. Building on this work, Lee et al. introduced techniques to prevent GPS spoofing for static receivers~\cite{gps-anti-spoofing-static}. And Semperfi et. al.~\cite{gps-anti-spoofing-semperfi} developed anti-spoofing technique for mobile GPS receivers, such as UAVs, enhancing robustness for diverse navigation systems. Location and time information in GPS signals is tightly coupled. It means that these techniques mitigate both location and time-sync errors under adversarial conditions.
\section{Theoretical Tools for Securing Time}
In this section, we discuss research work that employs theoretical tools to secure the timing stack. In this context, theoretical tools are employed for three distinct tasks: i) establish properties of a system model, ii) proving correctness of a system design i.e. it aligns with the stated goals and iii) verify software implementation of a given time-stack component. 

\subsection{Hardware}
The timing vulnerabilities in the hardware layer primarily result from either physical side channels ($I01$, $I02$) or design limitations ($I03$, $I04$).
Traditional formal verification tools cannot address these issues because mitigating them requires either the addition of new components or changing the existing designs. However, these new designs may be evaluated using standard formal verification tools~\cite{formal-verification-intel}.

\subsection{Software}
Timekeeping functionalities, integrated within broader system software like operating systems or hypervisors, are prone to inherent software vulnerabilities ($I05, I06, I07$). Although formal verification tools offer a means to analyze these systems for potential flaws, the large code bases and complex interaction among various subsystems of the system software make it infeasible to verify them. Despite these challenges, advancements in formal verification techniques have enabled the verification of specific OS components~\cite{formal-verification-eBPF} and hypervisors~\cite{formal-verification-hypervisor-arm, formal-verification-hypervisor-memory, formal-verification-kvm}. Employing these verification tools to assess the security aspects of system software can diminish the timekeeping software's vulnerability exposure ($I05, I06, I07$). Further, timing subsystems may also contain vulnerabilities originating from incorrect implementations. To enhance their security, it is crucial to apply the latest formal verification tools to verify the correctness of timing subsystems including trusted timing services such as Timeseal~\cite{time-stack-timeseal} and T3E~\cite{trusted-time-t3e}. As far as our knowledge extends, applying formal verification tools to timekeeping software remains an open area of research.

\subsection{Network} \label{subsec:network-theory}
The use of theoretical tool to establish trust in the timing stacks have almost exclusively focused on its network component i.e. time synchronization. This literature has focused on following lines of work:

\noindent\textbf{\texttt{T01.} Establishing Secure Time Synchronization Requirements.} 
Theorem proving tools have been used to establish requirements for secure clock synchronization. Narula et al.\cite{net-sync-gps-sec-transfer} constructed formal models for one-way and two-way time transfer, assuming a line-of-sight link between the systems and a threat model with a MITM adversary. They present proves for i) one-way time transfer's inherent susceptibility to delay attacks (as discussed in $I12$) and ii) essential requirements for a two-way secure time-synchronization protocol. Building upon this work, they study two-way time synchronization over a multi-hop network where systems at the both ends implement cryptography~\cite{net-sync-gps-sec-sync}. They put forward the prerequisites for a secure clock synchronization algorithm applicable to protocols like PTP~\cite{ptp-std-doc}. Among other requirements, they show that the timing packets must travel along the shortest path between the server and the client to completely prevent delay attacks. This has an important implication that \textit{delay attacks, over the network, cannot be prevented entirely if they do not guarantee shortest path traversal.} This is indeed the case of today's internet that employs TCP/IP stack for networking.

\begin{table*}[t]
\footnotesize
\centering
\begin{tabular}{p{2 cm} p{3.5cm}  p{0.5cm}  p{0.5cm}  p{0.5cm}  p{0.5cm}  p{0.5cm}  p{0.5cm}  p{0.5cm} p{0.5cm}  p{0.5cm}  p{0.5cm}  p{0.5cm}  p{0.5cm} }
 \multicolumn{1}{c}{} & \multicolumn{1}{c}{} & \multicolumn{8}{c}{Systems Approach} & \multicolumn{4}{c}{Theoretical Approach} \\
 \cmidrule(lr){3-10} \cmidrule(lr){11-14}
    & & \texttt{D01} & \texttt{D02} & \texttt{D03} & \texttt{D04} & \texttt{D05} & \texttt{D06} & \texttt{D07} & \texttt{D08} & \texttt{T01} & \texttt{T02} & \texttt{T03} & \texttt{T04} \\
 \hline
  Hardware Issues & Physical Side Channels & \halfcirc & \emptycirc & \emptycirc & \emptycirc & \emptycirc & \emptycirc & \emptycirc & \emptycirc & \emptycirc & \emptycirc & \emptycirc & \emptycirc \\
& Design Limitations & \emptycirc & \halfcirc & \emptycirc & \emptycirc & \emptycirc & \emptycirc & \emptycirc & \emptycirc & \emptycirc & \emptycirc & \emptycirc & \emptycirc \\
\hline
Software Issues & System Software Bugs & \emptycirc & \emptycirc & \halfcirc & \emptycirc & \emptycirc & \emptycirc & \emptycirc & \emptycirc & \emptycirc & \emptycirc & \emptycirc & \emptycirc \\
& TEE Limitations & \emptycirc & \emptycirc & \halfcirc & \fullcirc & \emptycirc & \emptycirc & \emptycirc & \emptycirc & \emptycirc & \emptycirc & \emptycirc & \emptycirc \\
\hline
 & Limited Crypto Adoption & \emptycirc & \emptycirc & \emptycirc & \emptycirc & \halfcirc & \emptycirc & \emptycirc & \halfcirc & \fullcirc & \halfcirc & \emptycirc & \emptycirc \\
Network Issues & Availability Issues & \emptycirc & \emptycirc & \emptycirc & \emptycirc & \emptycirc & \halfcirc & \halfcirc & \halfcirc & \fullcirc & \emptycirc & \emptycirc & \halfcirc \\
& Implementation Issues & \emptycirc & \emptycirc & \emptycirc & \emptycirc & \emptycirc & \emptycirc & \emptycirc & \emptycirc & \emptycirc & \emptycirc & \halfcirc & \emptycirc \\
\hline
\end{tabular}
\caption{Research contributions towards mitigating timing stack issues utilize both \textit{system-based} and \textit{theoretical} approaches. A full circle indicates that the defense technique mitigates all issues within the category, a half circle suggests that some of the issues in a category are addressed, and an empty circle signifies the given defense's lack of mitigation for issues in the specified category. While system-based defenses tackle attack surfaces across all three layers of the timing stack, theoretical solutions predominantly concentrate on the network component.}
\label{tab:system-v-defense}
\end{table*}

\noindent\textbf{\texttt{T02.} Proving Correctness of the Time-Sync Protocols.} Formal verification tools have been used to prove the correctness of fault-tolerant clock sync protocols. For instance, Schwier et. al.~\cite{theory-MechanicalVO} used protocol verification system (PVS) to verify a generalized time-sync protocol's correctness based on conditions established by Schneider~\cite{theory-schneider-conditons} for byzantine faults. Improving on this work, Barsotti et al., ~\cite{theory-schneider, theory-deductivetools} used deductive tools to prove correctness of fault-tolerant clock synchronization algorithms proposed by Lamport-Melliar\cite{clock-sync-fault-presence} and Lundlies-Lynch~\cite{clock-sync-fault-tolerant}. This research offers a promising direction for formal verification of time-sync protocols dealing with malicious faults.

Recent efforts regarding secure time synchronization using mathematical analysis have shifted focus to wireless sensor networks~\cite{theory-attack-resilient-pulse-coupled, theory-self-stablizing}. Most of these works assume a MITM attack model where few nodes in the network are compromised ($I10-I15$). Wang et al.\cite{theory-attack-resilient-pulse-coupled} introduced an attack-resilient pulse-coupled synchronization scheme for wireless sensor networks, deriving necessary conditions and analytically proving that it guarantees secure synchronization in the presence of a single malicious node. Another work by Hoepman et. al.\cite{theory-self-stablizing} presented a self-stabilizing clock synchronization algorithm, for wireless sensor networks. Their design is secure against pulse delay attacks by malicious nodes and they providing proofs for the correctness of their random beacon scheduling algorithm. These works demonstrate the potential of using theoretical tools for verify time-sync protocol designs.

\noindent\textbf{\texttt{T03.} Verification of Protocol Implementations.} Formal verification techniques can be used to verify the implementations of time-sync protocols ($I17$). This is demonstrated by Luca et al., who performed the automated verification of the gossip time-sync protocol's~\cite{model-checking-gossip} implementation using parameterized model checking. However, gossip is rather a simple protocol, and the methods employed for its verification do not readily extend to more complex protocols such as NTP, PTP etc~\cite{net-sync-openchallenges}. However, despite its limitations, partial verification of time-sync implementations using the existing formal verification methods can yield important security insights. In one such instance, Dieter et al.\cite{theory-nts-specs} performed (partial) formal verification of Network Time Security (NTS -- RFC8915) specifications~\cite{nts-rfc} and discovered two vulnerabilities in the analyzed version~\cite{theory-nts-formal-analysis}, which are currently being addressed. It shows that future research on enabling complete verification of widely used time-sync protocols would greatly contribute to their security.

\noindent\textbf{\texttt{T04.} Mitigating Attacks on Time Synchronization.} Beyond analyzing complete protocol design, mathematical tools have also been used to study specific time-sync attacks. This research primarily focuses on mitigating delay attacks against time-sync protocols by a network adversary. For instance, Mizrahi et al. proposed a multi-path time synchronization scheme designed to resist delay attacks by a man-in-the-middle attacker. They leverage game theory to provide proofs for the delay resiliency of their design~\cite{multi-path-game-theory}. Similarly, Anto et al. propose modifications to PTP aimed at mitigating delay attacks and formally verified the correctness of their proposed updates~\cite{theory-formal-attack-1588}. Likewise, Moussa et al. proposed extensions to the PTP protocol and formally proved the correctness of these protocol extensions~\cite{theory-smart-grids, theory-ptp-extension}. However, the proposed extensions are domain specific as they rely on redundant master clocks on power grid substations mandated by IEC 61850. Another work by Lisova et al., took a game-theoretic approach, modeling the interaction of a man-in-the-middle attacker introducing asymmetric delays to PTP packets and a network inspection system collecting clock offset information. Their work uses a game-theoretic tool to predict attacker strategies and develop mitigation mechanisms accordingly~\cite{theory-game-theory-1588}. While delay attacks have been a dominant subject of this research, defenses against other attacks would equally benefit from the use of theoretical tools.
\section{Building a secure timing infrastructure}
Our evaluation of the existing time security research shows a patchwork of solutions aimed at mitigating one or few security issues. We advocate for a fundamentally different paradigm and put forward recommendations for designing an integrated trusted time stack.

\noindent\textbf{System-wide Trusted Timing Services.} We embrace \emph{hardware-software co-design} based approach for building an integrated trusted time stack. At the heart of this new time stack is a fixed frequency monotonic counter $\mathbf{C}$, which is unaffected by DVFS and immune to writes by the software. Access to this counter and its frequency must be an atomic operation and available system-wide irrespective of the software's privilege level, trust and virtualization status. System-wide availability allows each application or other software component to maintain its own \textit{local clock} in an untrusted environment ($I03-09$). More importantly, this design also mitigates $I09$ in contrast to existing solutions~\cite{time-stack-timeseal, trusted-time-t3e}. The second component to this design is secure time-synchronization, which should be achieved by a dedicated co-processor outside the control of untrusted software. TimeCard~\cite{time-card} by OCP Time Appliance Project~\cite{ocp-tap} offers inspiration for this design. While their design does not consider time-sync security in particular, there are no fundamental barriers in implementing it. Finally, such module must communicate the time-sync parameters to the applications securely. Again, we recommend introducing extra registers (with system-wide availability) on the SoC that are updated by the time-sync co-processor but cannot be written to by any processor controlled by the untrusted software. 

It is important to remember that providing unrestricted access to high resolution time to untrusted software is a double edged sword~\cite{browser-timing-side-channel} because despite immense benefits, it can also enable side channel attacks. Hence, we recommend enabling system firmware (e.g., SMM mode on x86, secure monitor on ARM, etc.) to control the resolution and access permissions to the counter $\mathbf{C}$. These firmwares afford better protection by dint of their small TCBs that can be formally analyzed. It allows us to provide flexibility to the system designers, who can configure trusted time stack to meet their needs, while ensuring reasonable amount of trust in time.

\noindent\textbf{Complimenting System Design with Theoretical Tools.} We recommend using theoretical tools in conjunction with the system-based approaches for verifying trusted time stack design and its implementations. They have made important contributions towards a secure time stack such as working out requirements for secure time-sync~\cite{net-sync-gps-sec-sync}, verifying implementations of time-sync protocols~\cite{model-checking-gossip}, proving the security guarantees offered by Chronos~\cite{net-sync-chronos} and identifying issues in the implementations of NTS~\cite{theory-nts-formal-analysis}. Yet, the use of these tools to improve time stack security has been rather limited as shown in table~\ref{tab:system-v-defense}. This is, in part, due to the lack of automated tools for time-sync protocol verification as pointed out by Swen et al~\cite{net-sync-openchallenges}. Development of an automated tool for verifying models with time and clock abstractions represents a key research challenge.

\noindent\textbf{Delay Attacks.} Delay attacks represent one of the most challenging problem for the timing stack. As discussed in $T01$, solving this problem for two-way time-sync requires network packets to traverse the shortest network path~\cite{net-sync-gps-sec-sync}, in addition to other mechanisms. An important implication is that these attacks cannot be avoided completely. Nevertheless, it is possible to deal with them as we discussed in $D06 \& D07$. However, these solutions either mitigate against limited adversaries ($D06$) or result in performance degradation ($D07$). Future research must investigate improving timing stack's resiliency to delay attacks with lower trade-offs than offered by the state-of-the-art.
\section{Conclusion}
We present a first systematization of the timing stack security. Our framework identifies three layers constituting a typical time stack: hardware, software and network. Through our framework, we identify new vulnerabilities of timing stack originating in the hardware and software layers. Further, we compare and contrast two classes i.e. system-based and theoretical of tools used for mitigating time stack security issues. We conclude by providing concrete recommendations for the design of a trusted timing stack.

\bibliographystyle{plain}
\bibliography{main}
\end{document}